\documentclass[12pt]{article}
\usepackage{epstopdf}
\usepackage{latexsym}
\usepackage{makeidx}
\makeindex
\usepackage{geometry}
\begin{document}
\pagestyle{headings}
\newcommand{\grad}{\mbox{\boldmath$\nabla$}}
\newcommand{\bdiv}{\mbox{\boldmath$\nabla\cdot$}}
\newcommand{\curl}{\mbox{\boldmath$\nabla\times$}}
\newcommand{\bcdot}{\mbox{\boldmath$\cdot$}}
\newcommand{\btimes}{\mbox{\boldmath$\times$}}
\newcommand{\btau}{\mbox{\boldmath$\tau$}}
\newcommand{\btheta}{\mbox{\boldmath$\theta$}}
\newcommand{\bmu}{\mbox{\boldmath$\mu$}}
\newcommand{\bepsilon}{\mbox{\boldmath$\epsilon$}}
\newcommand{\bcj}{\mbox{\boldmath$\cal J$}}
\newcommand{\bcf}{\mbox{\boldmath$\cal F$}}
\newcommand{\bbeta}{\mbox{\boldmath$\beta$}}
\newcommand{\lbar}{\lambda\hspace{-.09in}^-}
\newcommand{\bcp}{\mbox{\boldmath$\cal P$}}
\newcommand{\bco}{\mbox{\boldmath$\omega$}}
\newcommand{\brho}{\mbox{\boldmath$\rho$}}
\title{The nature of electromagnetic energy}
\author{Jerrold Franklin\footnote{Internet address:
Jerry.F@TEMPLE.EDU}\\
Department of Physics\\
Temple University, Philadelphia, PA 19122-6082
\date{\today}}
\maketitle
\begin{abstract}
We study the nature and location of electromagetic energy for two cases.
The energy density for electromagnetic radiation is shown to be
$\frac{1}{8\pi}(E^2+B^2)$, with the energy contained in the electromagnet fields.
For a static charge distribution, the electromagnet energy is contained in the charge,
with an energy density, $\frac{1}{2}\rho\phi$, There is no energy outside the charge distribution.  The electromagnetic fields do not contain the energy,
and $\frac{1}{8\pi}(E^2+B^2)$ cannot be considered an energy density in this case.
There is no ambiguity in either case as to where the energy is located.

\end{abstract}
\section{Introduction}
The concept of electromagnetic energy is generally introduced as a consequence of conservation of energy in the interaction of electromagnetic fields or potentials with charge-current distributions.
However, the form that electromagnetic energy takes, and its location  in space, has not been definitively settled.  This leads to the question of whether the electromagnetic fields in the form 
$\frac{1}{2}(E^2+B^2)$ or   charge distributions in the form $\frac{1}{2}\rho\phi$, should constitute electromagnetic energy densities.\footnote{A detailed history of this question is given in \cite{jr}, which includes many earlier references.}

For instance, Richard Feynman, in ``Lectures  on Physics''\cite{f}, wrote 
``The idea that [electrostatic] energy is located somewhere is not necessary'', and
 David Griffiths, in his textbook,  ``Introduction to Electrodynamics,  4th Edn.''\cite{dg}, asked the question,
 ``Where is the energy, then? Is it stored in the field, ...  or is it stored in the charge?  I can tell you what the total energy is,  and I can provide you with several different ways to compute it, but it is impertinent to worry about where the energy is located. In the context of radiation theory,
it is useful (and in general relativity it is essential) to regard the
 energy as stored in the field, with an energy density,
 $\frac{1}{2}(E^2+B^2)$.
 But in electrostatics one could just as well say it is stored in the charge, with a
 density, $\frac{1}{2}\rho\phi$.
 The difference is purely a matter of bookkeeping.''
 
 In the following sections of this paper, we resolve the question of when electromagnetic energy is stored in the electromagnetic fields and when it is in charge distributions.
It is not purely a matter of bookkeeping.
 
 \section{Poynting's theorem}

We start with the standard textbook derivation of Poynting's theorem, which introduces the concept of electromagnetic energy.

The rate at which an electric field gives energy to a charge $q$ moving with velocity $\bf v$ is 
$q{\bf v\bcdot E}$.
The magnetic field does not enter here because its force on the charge is always perpendicular to the velocity, and does no work.

The current density $\bf j$ is composed of many of these charges, such that
\begin{equation}
{\bf j}\Delta V=\sum_i q_i{\bf v_i},
\end{equation}
where the sum is over all the charges in the volume  $\Delta V$.

Although there are many charges within the volume $\Delta V$, it can be small enough that the macroscopic electric field is constant throughout the volume.
Then the rate at which  the electric field produces mechanical
energy into matter
within $\Delta V$ is given by
\begin{equation}
\frac{dU_{\rm matter}}{dt}=
\sum_i q_i{\bf v_i\bcdot E}={\bf j\bcdot E}\Delta V.
\end{equation}
It is clear here that the electric field acting on the charge is from other charges, so the concept of electromagnet self-energy doesn't arise.

If we sum over all these volumes in the limit as $\Delta V$$\rightarrow$0,
we get the integral relation
\begin{equation}
\frac{dU_{\rm matter}}{dt}=\int_V{\bf j\bcdot E}d^3r
\label{eq:utje}
\end{equation}
for the rate at which the electric field  puts mechanical
energy into matter .
By conservation of energy, the rate at which energy is put into the electromagetic fields is given by
\begin{equation}
\frac{dU_{\rm EM}}{dt}=-\frac{dU_{\rm matter}}{dt}
=-\int_V{\bf j\bcdot E}d^3r.
\label{eq:utj}
\end{equation}
  
We can replace the current $\bf j$ in Eq.\ (\ref{eq:utj}) by using Maxwell's
curl $\bf B $ equation, and
performing some vector manipulation.\footnote{We are using Gaussian units with c=1.}
Then
\begin{eqnarray}
\frac{dU_{\rm EM}}{dt} &=&
 \frac{1}{4\pi}\int_V[{\bf E}\bcdot \partial_t{\bf E}-{\bf E\bcdot(\curl B})]d^3r
\nonumber\\
&=&\frac{1}{4\pi}\int_V[{\bf E}\bcdot \partial_t{\bf E}-{\bf B\bcdot(\curl E)}-{\bf\bdiv(B\btimes E)}
]d^3r\nonumber\\
&=& \frac{1}{8\pi}\int_V\partial_t(E^2+B^2)d^3r+\frac{1}{4\pi}\oint_S{\bf dS\bcdot(E\btimes B)}. 
\label{jde2}
 \end{eqnarray}

We interpret each term in Eq.\ (\ref{jde2}) as follows:
\begin{displaymath}
\bullet\;\frac{dU_{\rm EM}}{dt}\, {\rm
\; is\; the\; rate\; at\; which\; energy\; is \; being\; put\; into\; the\; electromagnetic\; fields.}
\end{displaymath}
\begin{displaymath}
\bullet\; \frac{1}{8\pi}\int_V \partial_t(E^2+B^2)d^3r
{\rm\; is\; the\; rate\; of\; increase\; of\; the\; energy\; within; the\; volume\; V.}
\end{displaymath}

\begin{displaymath}
\hspace*{-.7in}\bullet\; \frac{1}{4\pi}\oint_S{\bf dS\bcdot(E\btimes B)}\;
{\rm\; is\; the\; rate\; at\; which\; electromagnetic\; energy}
\end{displaymath}
\vspace{-.2in}
\begin{displaymath}
{\rm leaves\; the\; volume\; through\; its\; surface.}
\end{displaymath}

Equation (\ref{jde2}) is called {\bf Poynting's Theorem},
and the vector in the surface integral is called the
{\bf Poynting vector},
\begin{equation}
{\bf S_P}=\frac{1}{4\pi}{\bf E\btimes B}.
\end{equation}
The Poynting vector has the significance that the rate of flow of electromagnetic energy through a differential surface {\bf dS} is given by
${\bf S_P\bcdot dS}$.

We see from Eq.~(\ref{jde2}) that the electromagnetic energy produced by matter in a volume, V,
goes to two places.
It  increases the energy in the $\bf E$ and $\bf B$ fields within the volume, 
while some of the energy is carried out of the volume by the Poynting vector.

\section{Energy in electromagnetic radiation}

Electromagnetic radiation in otherwise empty space is produced by oscillating charges and currents within a small volume, as given by Eq.~(\ref{eq:utje}).
The radiated energy is carried out of this volume by the Poynting vector.
The region outside the emitting volume contains no charges or currents, so Poynting's theorem for that volume becomes
\begin{eqnarray}
 \frac{1}{8\pi}\int_{V}\partial_t(E^2+B^2)d^3r&=&\frac{1}{4\pi}\oint_S{\bf dS\bcdot(E\btimes B)}. 
\label{eba}
\end{eqnarray}
where $V$, here, is the volume of any effectively empty space outside the region in which the energy is produced.

If Eq.~(\ref{eba}) is implemented starting with no electromagnetic energy within a volume, then the energy content of the electric and magnetic fields within the volume $V$ is given by
\begin{equation}
U_{\bf EB}=\frac{1}{8\pi}\int_V (E^2+B^2 )d^3r.
\label{eq:uem}
\end{equation}

  We can use 
Eq.\ (\ref{eq:uem}) to define an energy density of the electric and magnetic fields given by
\begin{equation}
u_{\bf EB}=\frac{1}{8\pi}(E^2+B^2).
\label{eq:dem}
\end{equation}
Then, the integral,
\begin{equation}
U_{\bf EB}=\frac{1}{8\pi}\int_V u_{\bf EB}d^3r.
\label{eq:uem2}
\end{equation}
gives the electromagnet energy within any voume, $V$.

\section{Energy in Charge distributions.}

We can use the definition of the electric field in terms of the scalar and vector potentials,
\begin{equation}
{\bf E}=-\grad\phi-\partial_t{\bf A},
\label{epa}
\end{equation}
to rewrite, Eq.~(\ref{eq:utj}) as 
\begin{eqnarray}
\frac{dU_{\rm EM}}{dt}
&=&\int_V{\bf j}\bcdot\left(\grad\phi+\partial_t{\bf A}\right)d^3r.
\end{eqnarray}

Integrating this equation with respect to time leads to
\begin{eqnarray}
U_{\rm EM}(t)=U_{\rho{\bf j}}(t)&=&
\int_0^t dt'\int_V{\bf j(t')}\bcdot\left(\grad\phi(t')+\partial_{t'}{\bf A(t')}\right)d^3r\nonumber\\
&=&\int_0^t dt'\int_V\left\{\bdiv[{\bf j(t')}\phi(t'))-\phi(\bdiv{\bf j(t')}]
+{\bf j(t')\bcdot(\partial_{t'} A(t')})\right\}d^3r\nonumber\\
&=&\int_0^t dt'\int{\bf dS\bcdot(j(t')}\phi(t'))
+\int_0^t dt'\int_V\left[\phi(\partial_{t'}\rho(t'))
+{\bf j\bcdot(\partial_{t'} A(t')})\right]d^3r.\\
\label{rhoj}
\end{eqnarray}
In the above,  we have used the continuity equation
to replace $\bdiv{\bf j(t')}$ by $-\partial_{t'}\rho(t')$.

To produce a static charge distribution we use a fixed current density, ${\bf j}(t')={\bf j_0}$ that ends at $t'=T$.
Then, the vector potential, $\bf A$, will be constant, with
\begin{eqnarray}
\partial_{t'}{\bf A}(t')&=&0.
\end{eqnarray}
The charge distribution and potential produce by the constant current will have  linear increases with time, so
\begin{eqnarray}
\rho(t')&=&t'\rho(t)/t,\\
\phi(t')&=&t'\phi(t)/t.
\end{eqnarray}
Then,
\begin{eqnarray}
U_{\rm EM}(t)&=&\int_0^t dt'\int{\bf dS\bcdot j_0}t'\phi(t))/t
+\int_0^t dt'\int_V\left[t'\rho(t)\phi(t)/t^2\right]d^3r\nonumber\\
&=&\int_V\frac{1}{2}\rho(t)\phi(t)d^3r
+\frac{1}{2}\int{\bf dS\bcdot j_0}\phi(t)).
\end{eqnarray}

The current ends at $t=T$, so $U_{\rm EM}, \rho$, and $\phi$ become constant, and
\begin{eqnarray}
U_{\rm EM}=U_{\rho\phi}
&=&\int_V\frac{1}{2}\rho\phi d^3r,\quad t>T.
\label{urp}
\end{eqnarray}
This shows that $\frac{1}{2}\rho\phi$ integrated over any volume gives the electromagnetic energy within the volume, so
\begin{eqnarray}
u_{\rho\phi}&=&\frac{1}{2}\rho\phi
\end{eqnarray}
is an energy density for the electromagnetic energy of a charge distribution. 

The energy of a charge distribution can be given in terms of the electric field by the steps,
\begin{eqnarray}
U_{\rm EM}&=&\int_V\frac{1}{2}\rho\phi d^3r\nonumber\\
&=&\frac{1}{8\pi}\int_V\phi{\bf(\bdiv E}) d^3r\nonumber\\
&=&\frac{1}{8\pi}\int_V[{\bf\bdiv(\phi E)-E\bcdot\grad\phi] }d^3r\nonumber\\
&=&\frac{1}{8\pi}\int_V E^2 d^3r
+\frac{1}{8\pi}\int{\bf dS\bcdot E}\phi.
\label{ue}
\end{eqnarray}
This shows that $\frac{1}{8\pi}E^2$ cannot be considered an energy density for the energy of a charge distribution here, because the surface integral in Eq.~(\ref{ue}) also contributes.
The volume integral of $\frac{1}{8\pi}E^2$ does give the energy in an infinite volume where the surface intregral vanishes, but does not itself give the energy in a finite volume. 

\section{Conclusion}

We have seen that the location of electromagnetic energy depends on the particular situation considered.

For electromagnetic radiation in empty space, Eq.~(\ref{eq:uem}) shows that
$u_{\bf EB}=\frac{1}{8\pi}(E^2+B^2)$ is an energy density whose integral over any volume gives the enery the energy within the volume.

For a static charge distribution, Eq.~(\ref{urp}) shows that 
$u_{\rho\phi}=\frac{1}{2}\rho\phi$ is an energy density whose integral over any volume gives the energy within the volume.  This means that the energy of a static charge distriution resides in the charge distribution, with no energy outside of the charge.  The integral of 
$u_{\bf EB}=\frac{1}{8\pi}E^2$ over all space gives the total energy of the charge,
but it does not give the energy within any finite volume, and cannot be considered an energy density for this case.

In each of the cases we considered, there is no  ambiguity about the location of the electromagnetic energy.  More complicated cases could be considered,  with the location of the energy depending on the particular case.

\end{document}